\documentclass{article}

\usepackage{microtype}
\usepackage{graphicx}
\usepackage{booktabs}
\usepackage{multirow}
\usepackage{array}
\newcommand{\yes}{$\bullet$}
\newcommand{\partly}{$\circ$}
\newcommand{\no}{--}
\usepackage{tikz}
\usetikzlibrary{arrows.meta,patterns}

\usepackage{hyperref}
\usepackage{xurl}

\usepackage[accepted]{mlsys2025}
\makeatletter
\renewcommand{\Notice@String}{Preprint. Under review.}
\makeatother

\mlsystitlerunning{PipeLink: Contract-Guided Linking of GPU Kernel Components}

\begin{document}

\twocolumn[
\mlsystitle{PipeLink: Contract-Guided Linking of GPU Kernel Components}

\begin{mlsysauthorlist}
\mlsysauthor{Jize Guo}{pku}
\mlsysauthor{Xupeng Miao}{pku}
\end{mlsysauthorlist}

\mlsysaffiliation{pku}{School of Computer Science, Peking University, Beijing, China}

\mlsyscorrespondingauthor{Xupeng Miao}{xupeng.miao@pku.edu.cn}

\mlsyskeywords{GPU kernels, kernel fusion, linking, megakernels, MLSys}

\vskip 0.3in

\begin{abstract}
The fastest machine-learning kernels are compositions of components, such as
loaders, matrix-multiply mainloops, epilogues and communication stages, that run
concurrently on one GPU streaming multiprocessor and share its on-chip memory,
barriers, warps and registers. Because these components hard-code where they
sit in this shared space, they cannot be compiled, shipped or replaced on their
own: combining components from different sources means forking them by hand or
keeping them in separate kernels. We present PipeLink, a contract-guided linker
for GPU kernel components. Each component is compiled separately into an LTO
object that carries a contract stating what it requires and what it promises;
PipeLink checks the contracts against each other, relocates every component by
solving one joint placement of the SM's resources, and generates the transfers
between them, and the vendor LTO backend folds the result into constants. Across
51 configurations on Hopper and Blackwell GPUs, linked kernels run within 0.2\%
of the same sources compiled as one unit. Linking makes new compositions
practical: a CuTe attention operator inside a ThunderKittens decode megakernel
speeds up decoding by 2.16$\times$, a four-owner MoE training megakernel matches
the original monolith, one unchanged attention core yields sparse and
context-parallel variants that outperform FlexAttention, vLLM and Transformer
Engine, and a GEMM + SwiGLU composition whose consumer serves three GPU
architectures runs 1.2$\times$ faster on Blackwell than cuBLAS with a compiled
activation.

\end{abstract}
]

\printAffiliationsAndNotice{}

\section{Introduction}
\label{sec:intro}

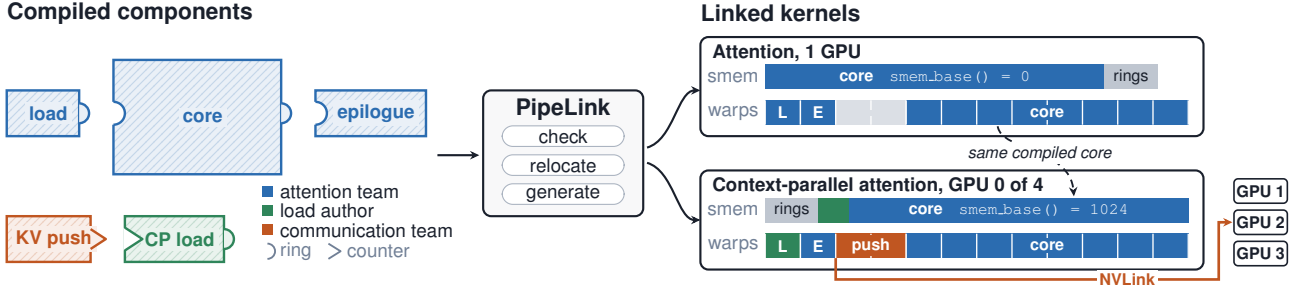
\begin{figure*}[t]
  \centering
  \resizebox{\textwidth}{!}{\def\fX{\fontsize{6}{7}\selectfont}\def\fS{\fontsize{7}{8.4}\selectfont}\def\fT{\fontsize{8.5}{10}\selectfont}
\begin{tikzpicture}[
  x=1cm, y=1cm, font=\sffamily\fS,
  >={Stealth[length=3.5pt,width=3.2pt]},
  ttl/.style={font=\sffamily\fT\bfseries, text=ink, anchor=west, inner sep=0},
  cell/.style={font=\sffamily\fX\bfseries, text=white, inner sep=0},
]
\definecolor{att}{HTML}{2B6CB0}   %
\definecolor{cpl}{HTML}{2F855A}   %
\definecolor{com}{HTML}{C05621}   %
\definecolor{ink}{HTML}{1A202C}
\definecolor{mute}{HTML}{718096}

\def\tile#1#2#3#4#5#6{%
  \fill[#5!10] (#1,#2) rectangle ++(#3,#4);
  \fill[pattern=north east lines, pattern color=#5!28] (#1,#2) rectangle ++(#3,#4);
  \draw[#5, line width=0.9pt] (#1,#2) rectangle ++(#3,#4);
  \node[font=\sffamily\fS\bfseries, text=#5, fill=#5!10, inner sep=1.5pt]
    at ({#1+(#3+0.14)/2},{#2+#4/2}) {#6};}
\def\rtab#1#2#3{%
  \fill[#3!10] (#1,{#2-0.13}) arc (-90:90:0.13) -- cycle;
  \draw[#3, line width=0.9pt] (#1,{#2-0.13}) arc (-90:90:0.13);}
\def\rsock#1#2#3{%
  \fill[white] ({#1-0.02},{#2+0.15}) -- (#1,{#2+0.15}) arc (90:-90:0.15) -- ({#1-0.02},{#2-0.15}) -- cycle;
  \draw[#3, line width=0.9pt] (#1,{#2+0.15}) arc (90:-90:0.15);}
\def\ttab#1#2#3{%
  \fill[#3!10] (#1,{#2-0.13}) -- ({#1+0.2},#2) -- (#1,{#2+0.13}) -- cycle;
  \draw[#3, line width=0.9pt, line join=round] (#1,{#2-0.13}) -- ({#1+0.2},#2) -- (#1,{#2+0.13});}
\def\tsock#1#2#3{%
  \fill[white] ({#1-0.02},{#2-0.15}) -- ({#1+0.23},#2) -- ({#1-0.02},{#2+0.15}) -- cycle;
  \draw[#3, line width=0.9pt, line join=round] (#1,{#2-0.15}) -- ({#1+0.23},#2) -- (#1,{#2+0.15});}

\node[ttl] at (0.2,3.62) {Compiled components};
\tile{0.2}{2.0}{0.95}{0.6}{att}{load}
\tile{1.6}{1.5}{2.2}{1.5}{att}{core}
\tile{4.25}{2.0}{1.45}{0.6}{att}{epilogue}
\tile{0.2}{0.35}{1.1}{0.6}{com}{KV push}
\tile{1.75}{0.35}{1.3}{0.6}{cpl}{CP load}
\rtab{1.15}{2.3}{att}  \rsock{1.6}{2.3}{att}
\rtab{3.8}{2.3}{att}   \rsock{4.25}{2.3}{att}
\ttab{1.3}{0.65}{com}  \tsock{1.75}{0.65}{cpl}
\rtab{3.05}{0.65}{cpl}

\foreach \y/\c/\t in {1.27/att/attention team, 1.02/cpl/load author, 0.77/com/communication team}
  { \fill[\c] (3.55,{\y-0.07}) rectangle ++(0.14,0.14);
    \node[anchor=west, text=ink, inner sep=0] at (3.78,\y) {\t}; }
\draw[mute, line width=0.7pt] (3.62,0.39) arc (-90:90:0.1);
\node[anchor=west, text=mute, inner sep=0] at (3.78,0.49) {ring};
\draw[mute, line width=0.7pt, line join=round] (4.42,0.39) -- ({4.42+0.16},0.49) -- (4.42,0.59);
\node[anchor=west, text=mute, inner sep=0] at (4.66,0.49) {counter};

\draw[->, ink, line width=0.8pt] (5.85,1.75) -- (6.4,1.75);
\draw[ink, line width=0.8pt, rounded corners=3pt, fill=ink!3] (6.45,0.95) rectangle (8.55,2.6);
\node[font=\sffamily\fT\bfseries, text=ink, inner sep=0] at (7.5,2.36) {PipeLink};
\foreach \y/\t in {2.0/check, 1.62/relocate, 1.24/generate}
  \node[draw=mute!60, fill=white, rounded corners=4pt, minimum width=1.6cm, minimum height=0.27cm,
        inner sep=0, text=ink] at (7.5,\y) {\t};
\draw[->, ink, line width=0.8pt] (8.6,1.85) .. controls (8.95,1.95) and (9.0,2.6) .. (9.3,2.6);
\draw[->, ink, line width=0.8pt] (8.6,1.65) .. controls (8.95,1.55) and (9.0,0.9) .. (9.3,0.9);

\node[ttl] at (9.3,3.62) {Linked kernels};
\def\cw{0.4625}
\def\kernel#1#2{%
  \draw[ink, line width=0.8pt, rounded corners=3pt] (9.3,#1) rectangle (15.9,{#1+1.3});
  \node[anchor=west, font=\sffamily\fS\bfseries, text=ink, inner sep=0] at (9.45,{#1+1.1}) {#2};
  \node[anchor=east, text=mute, inner sep=0] at (10.07,{#1+0.31}) {warps};
  \node[anchor=east, text=mute, inner sep=0] at (10.07,{#1+0.78}) {smem};}
\def\warp#1#2#3#4#5{%
  \fill[#4] ({10.15+#2*\cw},{#1+0.15}) rectangle ++({#3*\cw},0.32);
  \foreach \i in {0,...,#3} \draw[white, line width=0.5pt] ({10.15+(#2+\i)*\cw},{#1+0.15}) -- ++(0,0.32);
  \node[cell, fill=#4, inner sep=1pt] at ({10.15+(#2+#3/2)*\cw},{#1+0.31}) {#5};}

\kernel{2.0}{Attention, 1 GPU}
\warp{2.0}{0}{1}{att}{L}
\warp{2.0}{1}{1}{att}{E}
\warp{2.0}{2}{2}{mute!25}{}
\warp{2.0}{4}{8}{att}{core}
\fill[att] (10.15,2.62) rectangle ++(4.45,0.32);
\node[cell] at (12.375,2.78) {core\quad\ttfamily\mdseries smem\_base() = 0};
\fill[mute!45] (14.6,2.62) rectangle ++(0.7,0.32);
\node[cell, text=ink, font=\sffamily\fX] at (14.95,2.78) {rings};

\kernel{0.25}{Context-parallel attention, GPU 0 of 4}
\warp{0.25}{0}{1}{cpl}{L}
\warp{0.25}{1}{1}{att}{E}
\warp{0.25}{2}{2}{com}{push}
\warp{0.25}{4}{8}{att}{core}
\fill[mute!45] (10.15,0.87) rectangle ++(0.7,0.32);
\node[cell, text=ink, font=\sffamily\fX] at (10.5,1.03) {rings};
\fill[cpl] (10.85,0.87) rectangle ++(0.4,0.32);
\fill[att] (11.25,0.87) rectangle ++(4.45,0.32);
\node[cell] at (13.475,1.03) {core\quad\ttfamily\mdseries smem\_base() = 1024};

\draw[ink, densely dashed, line width=0.7pt, ->] (13.2,2.13) .. controls (13.25,1.8) and (14.15,1.85) .. (14.2,1.21);
\node[fill=white, text=ink, font=\sffamily\fX\itshape, inner sep=1pt] at (13.75,1.77) {same compiled core};

\foreach \g/\y in {1/1.12,2/0.71,3/0.30}
  { \draw[ink, line width=0.6pt, rounded corners=1.5pt] (16.3,\y) rectangle ++(0.7,0.32);
    \node[font=\sffamily\fX\bfseries, text=ink, inner sep=0] at (16.65,{\y+0.16}) {GPU \g}; }
\draw[com, line width=1pt, ->] (11.075,0.4) -- (11.075,0.12) -- (16.1,0.12) -- (16.1,0.87) -- (16.28,0.87);
\node[font=\sffamily\fX\bfseries, text=com, fill=white, inner sep=1pt] at (14.9,0.12) {NVLink};
\end{tikzpicture}}
  \caption{PipeLink overview. Separately compiled components carry contracts at
    their interfaces (left). PipeLink checks the contracts, relocates the
    components and generates the code between them, so the same compiled
    attention core lands at different shared-memory offsets in two kernels
    (right). Lane widths are schematic.}
  \label{fig:overview}
\end{figure*}

The fastest machine-learning kernels on modern GPUs are no longer single
operators. An attention kernel in the FlashAttention family runs a tile loader,
a matrix-multiply mainloop, a softmax stage and an epilogue as groups of warps
that execute concurrently and hand tiles to each other through on-chip memory~\cite{shah2024flashattention,zadouri2026flashattention}.
Megakernels go further and run an entire decode step or training pass inside
one persistent kernel~\cite{spector2025megakernel,cheng2025mirage,jin2026event,sul2026mixture}, and communication-overlapped kernels move data between
GPUs while they compute~\cite{chang2024flux,zheng2025tilelink,sul2025parallelkittens}.
Each such kernel is a composition of \emph{components} that cooperate at fine
grain on the same streaming multiprocessor (SM).

Yet these components cannot be built, shipped or replaced as components. They
agree on physical facts that no interface records: which bytes of shared memory
each one owns, which barriers and warps it uses, how many registers it may hold,
and which tile sizes its neighbors schedule against. Whoever wants to combine
components from different sources therefore faces a dilemma: fork the sources
and re-derive these facts by hand, or keep the components in separate kernels.
Megakernels take the first path: they are written as one translation unit, by
one team, in one library, so adopting a faster operator from another library,
such as a CuTe~\cite{thakkar2023cutlass} attention in a ThunderKittens~\cite{spector2025thunderkittens} megakernel~\cite{spector2025megakernel}, means porting it to the
megakernel's conventions by hand. New attention variants also start as forks of
an expert kernel, and systems that move context-parallel communication into
the kernel write or modify their own attention kernel~\cite{sul2025parallelkittens,qiang2026syncopate}.
Production context-parallel attention takes the second path: it calls an
unmodified FlashAttention kernel once per key/value block and runs the
transfers between GPUs and the correction of partial outputs as separate
operations around it~\cite{nvidia2026transformer,zhuzilin2026ring,fang2024usp},
paying for every kernel boundary in launches, memory traffic and lost overlap.

Existing tools compose kernels either before or below this boundary. Template
libraries, domain-specific languages and warp-specialization compilers compose
operations within one source program that they own~\cite{thakkar2023cutlass,dong2025flexattention,wang2026tilelang,chen2026tawa,soi2025optimal,yadav2025task}.
Device linking with link-time optimization (LTO) does combine separately
compiled code into one kernel at no runtime cost~\cite{sundaram2023nvjitlink},
but it links \emph{by symbol}: it records which names each fragment defines and
needs, and nothing about the resources it occupies or the facts its neighbors
compiled against. When one owner of a mixture-of-experts (MoE) training kernel
changes a tile size and recompiles alone, device linking accepts the stale
combination, and the kernel hangs or silently leaves outputs unwritten
(Section~\ref{sec:symbol-linking}).

We argue that kernel components need what object files gave whole programs: a
\emph{contract} that states what crosses the component boundary, and a linker
that \emph{relocates} each component into the kernel it joins. Today's
components are, in effect, absolute-addressed: they hard-code their positions in
a small on-chip address space that every other component shares. A relocatable
component instead declares what it requires and what it promises, and reads its
placement through bindings fixed only at link time. Our thesis is that such
contracts, checked and bound at link time, let separately compiled components
share one SM as efficiently as if their sources had been merged by hand.

This paper presents \emph{PipeLink}, a contract-guided linker for GPU kernel
components. Each component is compiled separately into an LTO object that
carries its contract. PipeLink checks the contracts against each other, solves
one joint placement of all components' on-chip resources for the composition at
hand, generates the code that moves tensors between components, and binds the
memory that communication components share across GPUs. The vendor LTO backend
then folds the placement into constants, so linking costs nothing at run time.

We evaluate PipeLink on Hopper and two generations of Blackwell GPUs. Across 51
timed configurations, linked kernels run within 0.2\% of the same sources
compiled as a single unit, and compositions that previously required a fork
become practical:
\begin{itemize}
  \item a CuTe attention operator inside a ThunderKittens Llama-1B decode
    megakernel decodes 2.16$\times$ faster at a 4096-token context;
  \item an MoE training megakernel split among four owners matches the original
    monolith (0.99--1.01$\times$) and survives every stale-owner upgrade;
  \item one unchanged attention core yields sparse attention up to 38$\times$
    faster than FlexAttention and context-parallel attention 1.20--1.31$\times$
    faster than Transformer Engine;
  \item a fused GEMM and SwiGLU for Qwen3-32B, whose consumer links with
    producers on three GPU architectures, runs 1.20$\times$ faster on B300 than
    cuBLAS followed by a compiled SwiGLU.
\end{itemize}

In summary, this paper makes the following contributions:
\begin{itemize}
  \item \textbf{Component contracts}, which state what a concurrently running
    GPU kernel component requires and promises, so that it can be compiled,
    distributed and replaced on its own, without its source
    (Section~\ref{sec:contracts}).
  \item \textbf{A contract-guided linker} that checks contracts against each
    other, relocates components by jointly solving their placement, and
    generates the data transfers between them (Section~\ref{sec:linking}).
  \item \textbf{Zero-cost composition of real ML kernels} across libraries,
    owners and GPU architectures (Section~\ref{sec:eval}).
\end{itemize}

\section{Motivation}
\label{sec:motivation}

\subsection{Anatomy of a Fused Kernel}
\label{sec:anatomy}

Our running example is context-parallel attention on Hopper
(Figure~\ref{fig:overview}). Its core is a warp-specialized attention kernel
derived from FlashAttention-3~\cite{shah2024flashattention}:
a load warp copies query, key and value tiles into shared memory, two consumer
warpgroups compute attention with an online softmax, and an epilogue warp
writes the output. In the context-parallel kernel, two more warps push the
local key/value shard to the other GPUs. All twelve warps run concurrently and
hand work to each other through mbarrier rings, named barriers and, across
GPUs, readiness counters. They share the SM's shared memory, barriers, warps
and register file, and how much of each a component may use depends on which
other components share the thread block.

\subsection{Components Are Absolute-Addressed}
\label{sec:absolute}

Fused-kernel code today fixes all of these assignments in its source. A
FlashAttention~\cite{shah2024flashattention} or CUTLASS~\cite{thakkar2023cutlass} kernel declares one \texttt{SharedStorage} struct
that lays out every buffer and barrier at a constant offset, numbers its named
barriers in an enumeration, selects roles by comparing the warp index against
constants, and calls \texttt{setmaxnreg} with constant budgets.
ThunderKittens~\cite{spector2025thunderkittens} megakernels~\cite{spector2025megakernel} likewise compile a fixed page and semaphore layout
into every operator.
In the vocabulary of linkers, these components are \emph{absolute-addressed}:
they hard-code their positions in an address space that is small, shared and
spread over several resource kinds.

Absolute addressing is harmless as long as one team owns the whole kernel. It
becomes the obstacle as soon as a component is meant to be reused. Components
written independently each start at shared-memory offset~0, barrier ID~1 and
warp~0, so combining them unchanged collides by construction; someone must
re-derive every position with knowledge of all the others.

\subsection{Fork or Split}
\label{sec:fork-split}

Because components cannot be relocated, combining components from different
sources leaves two routes: fork their sources into one kernel, or keep them in
separate kernels.

\paragraph{Fork.} Forking merges the sources into one translation unit and
re-derives every shared assignment by hand. Megakernels take this route: one
team owns an entire model step~\cite{spector2025megakernel,cheng2025mirage,jin2026event,sul2026mixture,zheng2026uniep},
and adopting an operator from another library means porting it to the
megakernel's page, semaphore and instruction conventions; recent work automates
the port with coding agents that write a new megakernel per model~\cite{li2026forgemegakernel,jaber2026automegakernel}.
New attention variants likewise start as forks of an expert kernel such as
FlashAttention~\cite{shah2024flashattention,zadouri2026flashattention}, and
systems that move context-parallel communication into the kernel write or
modify their own attention kernel~\cite{sul2025parallelkittens,qiang2026syncopate,hu2026cuco}.
Every fork is a new kernel to maintain.

\paragraph{Split.} Splitting keeps each component unmodified in its own kernel
and connects the kernels through global memory. Production context-parallel
attention takes this route: it calls an unmodified FlashAttention kernel once
per key/value block, moves blocks between GPUs with collective operations on
other streams, and corrects the partial outputs and log-sum-exp values in
separate operations~\cite{nvidia2026transformer,zhuzilin2026ring,fang2024usp,yang2025context}.
Splitting preserves reuse but pays for every boundary with launches, memory
round trips and lost overlap.

Template libraries avoid both routes only for the variants their authors
anticipated, such as masks and score functions over a generic attention
kernel~\cite{dong2025flexattention,ye2025flashinfer}.

\subsection{Symbol Linking Is Not Enough}
\label{sec:symbol-linking}

CUDA already supports separate compilation: relocatable device code and
link-time optimization (LTO) combine independently compiled device functions
into one kernel as fast as a single translation unit~\cite{scudiero2014separate,sundaram2023nvjitlink},
and NVIDIA's own libraries ship LTO fragments for this purpose~\cite{nvidia2026cufftlto,nvidia2026cufftdx}.
But device linking resolves \emph{symbols}: it records which names a fragment
defines and needs, and nothing about the warps, shared memory, barriers or
registers it occupies, or about the facts its neighbors compiled against.

\begin{figure}[t]
\centering
\def\fX{\fontsize{6}{7}\selectfont}\def\fS{\fontsize{7}{8.4}\selectfont}\def\fT{\fontsize{8.5}{10}\selectfont}%
\begin{tikzpicture}[
  x=1cm, y=1cm, font=\sffamily\fS,
  >={Stealth[length=3.5pt,width=3.2pt]},
  fact/.style={font=\ttfamily\fX, text=ink, inner sep=1pt, align=left},
  arr/.style={font=\sffamily\fX\itshape, text=mute, inner sep=1pt},
  ttl/.style={font=\sffamily\fS\bfseries, text=ink, anchor=base west, inner sep=0},
]
\definecolor{att}{HTML}{2B6CB0}   %
\definecolor{cpl}{HTML}{2F855A}   %
\definecolor{com}{HTML}{C05621}   %
\definecolor{fwk}{HTML}{6B46C1}   %
\definecolor{ink}{HTML}{1A202C}
\definecolor{mute}{HTML}{718096}

\def\owner#1#2#3#4#5{%
  \fill[#4!10] (#1,#2) rectangle ++(#3,0.5);
  \draw[#4, line width=0.9pt] (#1,#2) rectangle ++(#3,0.5);
  \node[font=\sffamily\fS\bfseries, text=#4, inner sep=0] at ({#1+#3/2},{#2+0.25}) {#5};}
\def\chip#1#2#3#4#5#6{%
  \ifnum#6=1
    \fill[#4, rounded corners=1.5pt] (#1,{#2-0.15}) rectangle ++(#3,0.3);
    \node[font=\sffamily\fX\bfseries, text=white, inner sep=0] at ({#1+#3/2},#2) {#5};
  \else
    \fill[#4!10, rounded corners=1.5pt] (#1,{#2-0.15}) rectangle ++(#3,0.3);
    \draw[#4, line width=0.7pt, rounded corners=1.5pt] (#1,{#2-0.15}) rectangle ++(#3,0.3);
    \node[font=\sffamily\fX\bfseries, text=#4, inner sep=0] at ({#1+#3/2},#2) {#5};
  \fi}
\def\tick#1#2{\draw[ink, line width=0.8pt, line cap=round, line join=round]
  ({#1-0.06},#2) -- ({#1-0.015},{#2-0.05}) -- ({#1+0.07},{#2+0.07});}
\def\cross#1#2{\draw[ink, line width=0.8pt, line cap=round]
  ({#1-0.055},{#2-0.055}) -- ({#1+0.055},{#2+0.055}) ({#1-0.055},{#2+0.055}) -- ({#1+0.055},{#2-0.055});}

\owner{2.6}{3.05}{3.0}{fwk}{framework}
\owner{0}{1.75}{1.85}{com}{communication}
\owner{3.25}{1.75}{1.7}{att}{GEMM}
\owner{6.35}{1.75}{1.85}{cpl}{activation}

\draw[->, ink, line width=0.7pt] (1.3,2.25) |- (2.55,3.3);
\node[fact, anchor=west] at (1.36,2.72) {DISPATCH\_Nb\\COMBINE\_Nb};
\draw[->, ink, line width=0.7pt] (1.9,2.07) -- (3.2,2.07);
\node[fact, anchor=north] at (2.55,2.02) {DISPATCH\_Nb};
\draw[->, ink, line width=0.7pt] (7.27,2.25) |- (5.65,3.3);
\node[fact, anchor=east] at (7.2,2.72) {PIPE\_DEPTH};

\draw[ink, line width=0.7pt, rounded corners=2pt, fill=mute!12] (2.75,0.86) rectangle (5.45,1.18);
\node[font=\ttfamily\fX, text=ink, inner sep=0] at (4.1,1.02) {routed\_buffers\_done};
\draw[->, ink, line width=0.7pt] (0.6,1.75) |- (2.7,1.02);
\draw[->, ink, line width=0.7pt] (4.1,1.75) -- (4.1,1.23);
\draw[->, ink, line width=0.7pt] (7.6,1.75) |- (5.5,1.02);
\node[arr, anchor=south] at (1.65,1.04) {arrive};
\node[arr, anchor=west] at (4.15,1.47) {arrive};
\node[arr, anchor=south] at (6.55,1.04) {arrive};

\draw[mute!50, line width=0.5pt] (0,0.6) -- (8.2,0.6);
\node[font=\sffamily\fS, text=ink, anchor=base west, inner sep=0] at (0,0.3)
  {\textbf{Upgrade:} communication repacks alone, \texttt{\fX DISPATCH\_Nb} 512 $\to$ 256};

\def\panel#1#2{%
  \draw[mute!70, line width=0.6pt, rounded corners=3pt] (#1,-1.45) rectangle ++(4.0,1.55);
  \node[ttl] at ({#1+0.12},-0.18) {#2};
  \chip{#1+0.12}{-0.95}{1.1}{com}{comm 256}{1}}

\panel{0}{linked by symbol}
\chip{2.18}{-0.7}{1.7}{fwk}{framework 512}{0}
\chip{2.18}{-1.2}{1.7}{att}{GEMM 512}{0}
\draw[mute, densely dashed, line width=0.6pt] (1.27,-0.95) -- (2.13,-0.72);
\draw[mute, densely dashed, line width=0.6pt] (1.27,-0.95) -- (2.13,-1.18);
\node[arr, anchor=south] at (0.75,-0.76) {symbols only};
\node[font=\sffamily\fX\bfseries, text=ink, anchor=base east, inner sep=0] at (3.62,-0.18) {hangs};
\cross{3.77}{-0.135}

\panel{4.2}{linked by contract}
\chip{6.38}{-0.7}{1.7}{fwk}{framework 256}{1}
\chip{6.38}{-1.2}{1.7}{att}{GEMM 256}{1}
\draw[->, ink, line width=0.7pt] (5.47,-0.95) -- (6.33,-0.72);
\draw[->, ink, line width=0.7pt] (5.47,-0.95) -- (6.33,-1.18);
\node[arr, text=ink, anchor=south] at (4.97,-0.76) {fact propagated};
\node[font=\sffamily\fX\bfseries, text=ink, anchor=base east, inner sep=0] at (7.82,-0.18) {passes};
\tick{7.98}{-0.135}
\end{tikzpicture}
\caption{Facts that symbol linking does not record, in the four-owner MoE
megakernel. Top: constants that one owner compiles and another schedules
against, and a counter shared by three owners. Bottom: the communication owner
halves its dispatch tile; linked by symbol the kernel hangs, linked by contract
it passes.}
\label{fig:stale-fact}
\end{figure}
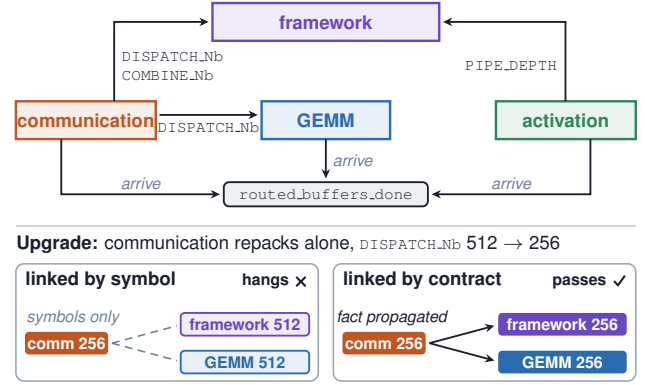

The consequence is silent failure. The four owners of our MoE megakernel
(Section~\ref{sec:eval-moe}) schedule against each other's tile geometry
(Figure~\ref{fig:stale-fact}). We changed one geometry constant of one owner at
a time, repacked only that owner, and linked it with the others' unchanged
objects. Device linking accepted all five combinations: three kernels hang, and
two complete without error but leave part of their output unwritten. Separately
compiled components need an interface that carries what they share, and a
linker that acts on it.

\section{Component Contracts}
\label{sec:contracts}

\subsection{Components}
\label{sec:components}

A \emph{component} is a separately compiled part of a fused kernel. PipeLink
supports three kinds. A \emph{role} is a group of warps with its own entry
function, such as a TMA loader, an MMA mainloop or a communication stage. A
\emph{callable} is a device function that runs on its caller's threads, such
as an attention mask, a score modifier or an operator body inside a
megakernel interpreter. A \emph{tensor
producer or consumer} is a role or callable that exports or imports a tile
held in on-chip storage. One source file may define several of these, and a
component may be written in plain CUDA, CUTLASS/CuTe, ThunderKittens or
TileLang.

The author packs each component into a \emph{package}: an LTO object produced
by the vendor compiler, plus a manifest that records its contract. Packing
compiles the source twice, once to evaluate the annotations that make up the
contract and once to produce the object. PipeLink never reads or rewrites the
component's instructions; to the linker the object is opaque, and the
component can be distributed without its source. The vendor LTO backend, which
does see the object's IR, later optimizes the composed kernel as a whole.

We considered composing at two other levels. Composing source or compiler IR,
as template libraries and warp-specialization compilers do, gives the composer
full freedom but requires every component's source in one compilation, under one
toolchain and one owner. Rewriting compiled machine code would avoid that, but
the GPU instruction set is undocumented and changes between generations, and
patched binaries miss the cross-component optimizations that make a fused
kernel fast. LTO objects sit between the two: they are a supported, stable
vendor format, they hide the source, and the backend still optimizes across
component boundaries.

\subsection{Contract Vocabulary}
\label{sec:vocabulary}

A contract states what a component \emph{requires} from the kernel it joins and
what it \emph{promises} to the other components (Table~\ref{tab:vocabulary}).

\begin{table}[t]
\caption{Contract vocabulary: what a component requires from the kernel it
joins, and what it promises to the other components.}
\label{tab:vocabulary}
\centering
\small
\begin{tabular}{@{}p{0.47\linewidth}p{0.47\linewidth}@{}}
\toprule
\textbf{Requires} & \textbf{Promises} \\
\midrule
Warps for each role & Roles and their entry functions \\
Shared memory, barriers, tensor memory, counters, with lifetimes & Channel endpoints and their protocols \\
Callables it calls & Callables it exports \\
Tensors it imports (demand) & Tensors it exports (placement) \\
Facts it schedules against & Facts it owns, e.g.\ tile sizes \\
Caller buffers and communication domains & Host parameters it builds \\
\bottomrule
\end{tabular}
\end{table}

Four properties of this vocabulary matter. First, every resource is claimed
by \emph{handle}, not by address: a claim says how much shared memory, how many
mbarrier slots or how many warps a component needs, never where they are.
Second, lifetimes let disjoint uses share storage: a claim may live for the
whole kernel, for one phase of a megakernel's instruction stream, or between
two events of a tensor connection. Third, synchronization
is described by protocol rather than by code: an mbarrier ring states its
arrival counts as functions of the participating roles' widths, its
transaction bytes, and which role initializes it, so that two endpoints can be
checked for agreement without inspecting either body. Inferring these
properties from the code instead would require analyzing opaque objects and
would still miss intent, such as which tile a buffer carries or when it may be
reused. Fourth, \emph{facts} are typed constants, such as a tile size, that one
component owns and others schedule against. A consumer does not compile a fact
in: it declares a kernel parameter sourced from the fact, and the linker checks
that all declarations agree and writes the owner's value into the kernel's
parameters, so a changed tile size (Figure~\ref{fig:stale-fact}) reaches every
consumer without recompiling it.

\subsection{Relocatable Components}
\label{sec:relocatable}

A component reads every link-time decision through a \emph{handle function}: an
external device function named after the handle, declared but not defined in
the component (Figure~\ref{fig:frontend}a, orange). Shared-memory bases, mbarrier slots,
named-barrier IDs, warp offsets, ring depths and counter slots are all read this
way. The object therefore contains unresolved calls instead of constants, much
as position-independent code reads addresses through a table filled in by the
dynamic linker. Unlike a table lookup, a handle function costs nothing at run
time: the linker defines it to return a constant that LTO folds away
(Section~\ref{sec:zero-cost}).

We need the object to stay unchanged across compositions, the binding to be
chosen per composition, and the result to cost nothing at run time; the obvious
alternatives each lose one of these.
Template or macro parameters fold into constants but require recompiling the
component for every composition, and so its source. Passing placements as
kernel arguments or through constant memory keeps the object fixed but turns
every offset into a run-time load and blocks constant folding in address
arithmetic. Relocation entries patched into the binary, as CPU linkers use, do
not exist for most of these resources: named-barrier IDs, warp ranges and
register budgets are instruction operands or control flow, not addresses.
Handle functions keep the object unchanged and still compile to immediates.

\subsection{Authoring}
\label{sec:authoring}

Annotations live in the component's own source file, next to the code they
describe, and are evaluated by the compiler at pack time. The author states
only what the compiler cannot see: semantic promises such as which tile an
export carries, which protocol a ring follows, which role initializes a
barrier, and how long a buffer stays live. Everything the compiler can observe
is captured instead of restated. Packing measures the size, alignment and field
offsets of parameter and descriptor types, records which device symbols the
object defines and needs, reads launch attributes from the compiled object,
and fits every enumerated layout map to a closed-form expression, rejecting the
package if the fit is not exact.

For tensor imports written in TileLang, the frontend captures the consumer's
demand from the layout of its destination fragment, so the author writes no
placement facts at all (Figure~\ref{fig:frontend}). Optionally, the frontend specializes the consumer to a
particular producer by adopting the producer's published register layout; the
result still links with any other producer, but the linker can then choose a
direct register copy instead of a staged transfer.

Figure~\ref{fig:frontend} compares the two SwiGLU consumers of our GEMM +
SwiGLU case study, which import the same accumulator. The CUDA consumer states
its contract in 10 device-side annotations spanning 38 lines; the TileLang
consumer writes no annotations and 14 lines of frontend calls, which name the
export to import and choose the chunk size, specialization and completion
policy. The rest is taken from the producer's contract or captured from the
compiled kernel. Both consumers share the same 9 host annotations and host
provider; the consumer-specific source is 163 lines in CUDA and 56 in TileLang.

Elsewhere in our case studies, a new attention load or communication component
takes 13--20 annotation statements and a megakernel operator 11--15; the four
owners of the MoE megakernel write 327 statements in total, against 3,482 lines
of component and host-provider source.

\begin{figure}[t]
\centering
\begin{tikzpicture}[x=1cm, y=1cm,
  ttl/.style={font=\sffamily\fontsize{7}{8.4}\selectfont\bfseries, text=fink, anchor=west, inner sep=0},
  code/.style={font=\ttfamily\fontsize{6}{7.2}\selectfont\frenchspacing, text=fink, anchor=west, inner sep=0},
  leg/.style={font=\sffamily\fontsize{6}{7.2}\selectfont, text=fink, anchor=west, inner sep=0}]
\definecolor{fprod}{HTML}{2B6CB0}
\definecolor{fcapt}{HTML}{2F855A}
\definecolor{fboth}{HTML}{718096}
\definecolor{fhandle}{HTML}{C05621}
\definecolor{fink}{HTML}{1A202C}
\colorlet{mute}{fboth}
\node[ttl] at (0,0.000) {(a) CUDA consumer: device-side contract and its use (abridged)};
\fill[fboth!12] (0.1,-0.491) rectangle (8.10,-0.229);
\fill[fboth] (0,-0.491) rectangle (0.08,-0.229);
\node[code] at (0.22,-0.360) {PL\_STAGE("swiglu.cuda",\ "1.0.0");};
\fill[fcapt!12] (0.1,-0.753) rectangle (8.10,-0.491);
\fill[fcapt] (0,-0.753) rectangle (0.08,-0.491);
\node[code] at (0.22,-0.622) {PL\_TYPE(tile\_args,\ a,\ b,\ M,\ N,\ K,\ m0,\ n0);};
\fill[fcapt!12] (0.1,-1.015) rectangle (8.10,-0.753);
\fill[fcapt] (0,-1.015) rectangle (0.08,-0.753);
\node[code] at (0.22,-0.884) {PL\_ROLE(epilogue,\ epilogue\_role,\ THREADS\ /\ 32);};
\fill[fprod!12] (0.1,-1.277) rectangle (8.10,-1.015);
\fill[fprod] (0,-1.277) rectangle (0.08,-1.015);
\node[code] at (0.22,-1.146) {PL\_CALL(tensor\_produce,\ ...,\ pl::produce,\ ...);};
\fill[fprod!12] (0.1,-1.539) rectangle (8.10,-1.277);
\fill[fprod] (0,-1.539) rectangle (0.08,-1.277);
\node[code] at (0.22,-1.408) {PL\_CALL(tensor\_release,\ ...,\ pl::produce,\ ...);};
\fill[fboth!12] (0.1,-1.801) rectangle (8.10,-1.539);
\fill[fboth] (0,-1.801) rectangle (0.08,-1.539);
\node[code] at (0.22,-1.670) {PL\_BINDING("chunk",\ pl::coord\{"chunk",\ 0,\ 8\},\ ...);};
\fill[fcapt!12] (0.1,-2.063) rectangle (8.10,-1.801);
\fill[fcapt] (0,-2.063) rectangle (0.08,-1.801);
\node[code] at (0.22,-1.932) {PL\_SITE(tensor\_import\_site,\ "epilogue",\ pl::participants\{};
\fill[fcapt!12] (0.1,-2.325) rectangle (8.10,-2.063);
\fill[fcapt] (0,-2.325) rectangle (0.08,-2.063);
\node[code] at (0.22,-2.194) {\ \ \ \ pl::coord\{"t",\ F,\ F\ +\ 256\},\ [](int\ t)\ \{\ return\ t;\ \}\},};
\fill[fcapt!12] (0.1,-2.587) rectangle (8.10,-2.325);
\fill[fcapt] (0,-2.587) rectangle (0.08,-2.325);
\node[code] at (0.22,-2.456) {\ \ "tile\_work",\ pl::guarantees\{"enter\_once",\ "converged",};
\fill[fcapt!12] (0.1,-2.849) rectangle (8.10,-2.587);
\fill[fcapt] (0,-2.849) rectangle (0.08,-2.587);
\node[code] at (0.22,-2.718) {\ \ \ \ "full\_warps",\ "no\_early\_return"\},\ ...);};
\fill[fcapt!12] (0.1,-3.111) rectangle (8.10,-2.849);
\fill[fcapt] (0,-3.111) rectangle (0.08,-2.849);
\node[code] at (0.22,-2.980) {PL\_CLAIM\_SMEM("output\_smem",\ OutputTile,\ ...);};
\fill[fboth!12] (0.1,-3.373) rectangle (8.10,-3.111);
\fill[fboth] (0,-3.373) rectangle (0.08,-3.111);
\node[code] at (0.22,-3.242) {PL\_IMPORT\_TENSOR(acc,\ import\_acc,};
\fill[fprod!12] (0.1,-3.635) rectangle (8.10,-3.373);
\fill[fprod] (0,-3.635) rectangle (0.08,-3.373);
\node[code] at (0.22,-3.504) {\ \ pl::identity\{"accumulator",\ "tile\_work",\ "completed\_k",};
\fill[fprod!12] (0.1,-3.897) rectangle (8.10,-3.635);
\fill[fprod] (0,-3.897) rectangle (0.08,-3.635);
\node[code] at (0.22,-3.766) {\ \ \ \ "f32",\ ...,\ \{\{"m",\ 0,\ 128\},\ \{"n",\ 0,\ 256\}\}\},};
\fill[fcapt!12] (0.1,-4.159) rectangle (8.10,-3.897);
\fill[fcapt] (0,-4.159) rectangle (0.08,-3.897);
\node[code] at (0.22,-4.028) {\ \ pl::demand\_positions\{pl::coord\{"t",\ 0,\ 256\},};
\fill[fcapt!12] (0.1,-4.421) rectangle (8.10,-4.159);
\fill[fcapt] (0,-4.421) rectangle (0.08,-4.159);
\node[code] at (0.22,-4.290) {\ \ \ \ pl::coord\{"p",\ 0,\ 16\},\ pl::bind("chunk"),};
\fill[fcapt!12] (0.1,-4.683) rectangle (8.10,-4.421);
\fill[fcapt] (0,-4.683) rectangle (0.08,-4.421);
\node[code] at (0.22,-4.552) {\ \ \ \ [](int\ t,\ int\ p,\ pl::bound\ c)\ \{};
\fill[fcapt!12] (0.1,-4.945) rectangle (8.10,-4.683);
\fill[fcapt] (0,-4.945) rectangle (0.08,-4.683);
\node[code] at (0.22,-4.814) {\ \ \ \ \ \ return\ \{row(t),\ 32*c\ +\ 16*half\_row(t)\ +\ p,\ ...\};\ \}\},};
\fill[fcapt!12] (0.1,-5.207) rectangle (8.10,-4.945);
\fill[fcapt] (0,-5.207) rectangle (0.08,-4.945);
\node[code] at (0.22,-5.076) {\ \ pl::lends\{pl::frame(...),\ pl::smem("scratch",\ ...),};
\fill[fcapt!12] (0.1,-5.469) rectangle (8.10,-5.207);
\fill[fcapt] (0,-5.469) rectangle (0.08,-5.207);
\node[code] at (0.22,-5.338) {\ \ \ \ pl::regs("acc\_dest",\ 16,\ ...),\ pl::cta\_barrier(...)\},};
\fill[fcapt!12] (0.1,-5.731) rectangle (8.10,-5.469);
\fill[fcapt] (0,-5.731) rectangle (0.08,-5.469);
\node[code] at (0.22,-5.600) {\ \ \{pl::rendezvous\{"tensor\_ready",\ ...\},};
\fill[fcapt!12] (0.1,-5.993) rectangle (8.10,-5.731);
\fill[fcapt] (0,-5.993) rectangle (0.08,-5.731);
\node[code] at (0.22,-5.862) {\ \ \ pl::rendezvous\{"import\_done",\ ...\}\});};
\node[code, text=mute] at (0.22,-6.124) {//\ kernel\ code\ uses\ the\ claim\ and\ the\ generated\ import:};
\fill[fhandle!12] (0.1,-6.517) rectangle (8.10,-6.255);
\fill[fhandle] (0,-6.517) rectangle (0.08,-6.255);
\node[code] at (0.22,-6.386) {auto\&\ tile\ =\ *(OutputTile*)(smem\ +\ output\_smem());};
\fill[fhandle!12] (0.1,-6.779) rectangle (8.10,-6.517);
\fill[fhandle] (0,-6.779) rectangle (0.08,-6.517);
\node[code] at (0.22,-6.648) {import\_acc(source,\ chunk,\ acc,\ scratch);};
\draw[mute!60, line width=0.4pt] (0,-0.229) rectangle (8.10,-6.779);
\node[ttl] at (0,-7.150) {(b) TileLang consumer: all it writes for the same contract};
\fill[fboth!12] (0.1,-7.641) rectangle (8.10,-7.379);
\fill[fboth] (0,-7.641) rectangle (0.08,-7.379);
\node[code] at (0.22,-7.510) {acc\ =\ pl.import\_tensor(partner,\ "accumulator",};
\fill[fboth!12] (0.1,-7.903) rectangle (8.10,-7.641);
\fill[fboth] (0,-7.903) rectangle (0.08,-7.641);
\node[code] at (0.22,-7.772) {\ \ \ \ \ \ \ \ \ \ \ \ \ \ \ \ \ \ \ \ \ \ \ outputs=("m0",\ "n0"),\ scratch\_bytes=0)};
\fill[fboth!12] (0.1,-8.165) rectangle (8.10,-7.903);
\fill[fboth] (0,-8.165) rectangle (0.08,-7.903);
\node[code] at (0.22,-8.034) {frag\ =\ acc.fragment(specialize=True,\ chunk=("n",\ 32))};
\fill[fboth!12] (0.1,-8.427) rectangle (8.10,-8.165);
\fill[fboth] (0,-8.427) rectangle (0.08,-8.165);
\node[code] at (0.22,-8.296) {with\ acc.produced(completion="final"):};
\fill[fhandle!12] (0.1,-8.689) rectangle (8.10,-8.427);
\fill[fhandle] (0,-8.689) rectangle (0.08,-8.427);
\node[code] at (0.22,-8.558) {\ \ \ \ staging\ =\ acc.staging((128,\ 128),\ "bfloat16")};
\node[code] at (0.22,-8.820) {\ \ \ \ for\ chunk\ in\ T.serial(8):};
\fill[fhandle!12] (0.1,-9.213) rectangle (8.10,-8.951);
\fill[fhandle] (0,-9.213) rectangle (0.08,-8.951);
\node[code] at (0.22,-9.082) {\ \ \ \ \ \ \ \ with\ acc.imported(frag,\ chunk=chunk):};
\node[code] at (0.22,-9.344) {\ \ \ \ \ \ \ \ \ \ \ \ ...\ \ \#\ SwiGLU\ on\ frag,\ then\ store\ staging};
\fill[fboth!12] (0.1,-9.737) rectangle (8.10,-9.475);
\fill[fboth] (0,-9.737) rectangle (0.08,-9.475);
\node[code] at (0.22,-9.606) {STAGE\ =\ pl.contract\_stage("swiglu.tilelang",};
\fill[fboth!12] (0.1,-9.999) rectangle (8.10,-9.737);
\fill[fboth] (0,-9.999) rectangle (0.08,-9.737);
\node[code] at (0.22,-9.868) {\ \ \ \ \ \ \ \ \ \ \ \ \ \ \ \ \ \ \ \ \ \ \ \ \ \ "1.0.0",\ program,\ host=...)};
\draw[mute!60, line width=0.4pt] (0,-7.379) rectangle (8.10,-9.999);
\fill[fprod] (0.00,-10.460) rectangle (0.08,-10.280);
\node[leg] at (0.16,-10.370) {from the producer's contract};
\fill[fcapt] (3.60,-10.460) rectangle (3.68,-10.280);
\node[leg] at (3.76,-10.370) {captured by the frontend};
\fill[fboth] (0.00,-10.760) rectangle (0.08,-10.580);
\node[leg] at (0.16,-10.670) {written in both};
\fill[fhandle] (3.60,-10.760) rectangle (3.68,-10.580);
\node[leg] at (3.76,-10.670) {uses a claim or the generated import};
\end{tikzpicture}
\caption{The same tensor-import contract written by hand in CUDA (a) and through
the TileLang frontend (b), abridged. Colors in (a) mark where the TileLang
version gets each fact instead. Orange lines use a claim and the generated
import; in TileLang the frontend emits the claim's handle read.}
\label{fig:frontend}
\end{figure}

\section{Contract-Guided Linking}
\label{sec:linking}

A \emph{pipeline manifest} lists the packages of one composition. Linking it
takes four steps (Figure~\ref{fig:overview}): check the contracts, solve the
placement, generate the connecting code, and device-link everything with the
vendor LTO backend. The output is one kernel image and a record of every
decision.

\subsection{Checking}
\label{sec:checking}

The linker pairs endpoints by name and checks that each pair agrees on memory
space, scope, tile, layout and protocol; for an mbarrier ring, it also checks
arrival counts, transaction bytes and which warp initializes the barrier.
Callables must match their signatures, facts must agree between their owner and
their consumers, and topologies such as the cluster shape are intersected
across components. Every rejection names one of 44 clauses together with the
conflicting facts: a system-scope counter producer paired with a grid-scope
consumer, for example, fails with \texttt{L-PAIR}, and a tensor-memory claim of
544 columns fails with \texttt{L-OVERLAP}.

Exports that nothing calls are dropped, as an ordinary linker drops
unreferenced sections, so one package can serve several compositions that each
call a different subset of its exports.

\subsection{Relocation as Constraint Solving}
\label{sec:relocation}

A classical linker relocates sections by concatenating them. That does not
work on an SM: its resource spaces are small (one thread block gets at most
227\,KiB of shared memory, 32 warps and 16 named barriers) and coupled, because the warps a role
occupies decide its register group, register groups share one register file,
and buffers may share storage only when their lifetimes are disjoint.

PipeLink instead solves one constraint problem per composition with an SMT
solver~\cite{demoura2008z3}.
The variables are the base of every claimed handle in every domain, the ring
depths left to the linker, and the register budget of every register group.
The constraints encode capacity, alignment, exact coverage of the warps by
roles, and non-overlap. Two claims may overlap only if their phases are
mutually exclusive or their lifetimes on one tensor connection are ordered, so
an epilogue's scratch can reuse a mainloop's operand ring once the accumulator
is ready. The objective prefers deeper rings and larger register budgets, then
a smaller footprint. The solver is deterministic, so a composition always
links to the same placement.

A greedy allocator would be simpler, but lifetimes turn placement into
packing, where the order of placement decides the footprint. In the MoE
megakernel, first-fit packing in lexical or decreasing-size order misses the
smallest shared-memory footprint, which the solver finds and which leaves
1,440 of 232,448 bytes free. Our compositions have 4--42 handles each; the
solver proves every objective optimal within 1.3\% of its work limit.

PipeLink also never reschedules a component. Warp-specialization compilers
choose pipelining and role assignment for the whole kernel, but they need the
source, and the owners of our components have already tuned their schedules for
their hardware. Keeping each schedule fixed makes the linked kernel's
performance predictable from its parts and keeps the linker out of the
components' correctness arguments.

The vendor backend, not the solver, allocates registers. After the device link,
PipeLink reads the linked kernel's register count and re-solves once if the
register groups do not fit; if they still do not fit, the link fails with
\texttt{L-REGS}.

\subsection{Why Linking Costs Nothing}
\label{sec:zero-cost}

The linker emits three pieces of code: \emph{handle stubs} that define every
handle function to return its solved constant, \emph{call trampolines} that
connect each call site to the export serving it, and a \emph{dispatch kernel}
that assigns warps to roles and sets each register group's budget once. All
three are compiled to LTO IR and device-linked with the components through
nvJitLink~\cite{sundaram2023nvjitlink}.
The backend inlines the stubs and folds their constants into the components'
address arithmetic, so a call such as \texttt{smem\_base()} becomes the
immediate \texttt{1024}. The result is the code a programmer would get by
writing the components into one source file with the same placement.
Dispatching through function pointers would be simpler to generate, but
indirect calls are not inlined, so constants would not fold and every role
would run at the register budget of the most demanding one. Section~\ref{sec:eval-setup}
compares the machine code of linked and single-unit kernels.

\subsection{Generated Tensor Transfers}
\label{sec:transfers}

A GEMM mainloop leaves its accumulator in tensor memory or in registers, and an
epilogue wants each of its threads to hold particular elements. Writing that
transfer by hand requires knowing the producer's layout, which is the coupling
contracts are meant to remove.

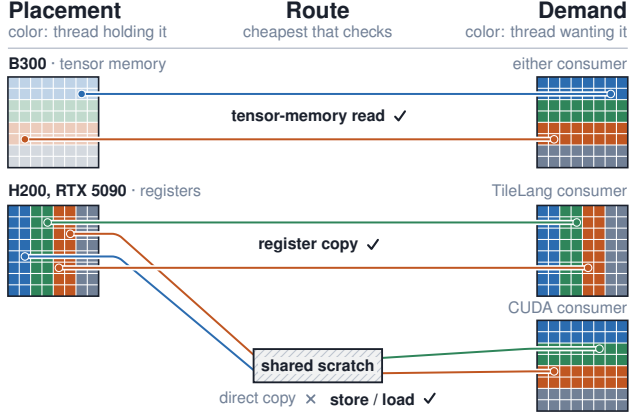
\begin{figure}[t]
\centering
\def\fX{\fontsize{6}{7}\selectfont}\def\fS{\fontsize{7}{8.4}\selectfont}\def\fT{\fontsize{8.5}{10}\selectfont}%
\begin{tikzpicture}[
  x=1cm, y=1cm, font=\sffamily\fS,
  >={Stealth[length=3.5pt,width=3.2pt]},
  hd/.style={font=\sffamily\fT\bfseries, text=ink, inner sep=0},
  sub/.style={font=\sffamily\fX, text=mute, inner sep=0},
  plat/.style={font=\sffamily\fX\bfseries, text=ink, anchor=base west, inner sep=0},
  note/.style={font=\sffamily\fX, text=mute, inner sep=0},
  rt/.style={font=\sffamily\fX\bfseries, text=ink, fill=white, inner sep=1.2pt},
]
\definecolor{att}{HTML}{2B6CB0}
\definecolor{cpl}{HTML}{2F855A}
\definecolor{com}{HTML}{C05621}
\definecolor{ink}{HTML}{1A202C}
\definecolor{mute}{HTML}{718096}

\def\own#1{\ifcase#1 att\or cpl\or com\or mute\fi}
\def\cs{0.15}%
\def\grid#1#2#3#4{%
  \foreach \r in {0,...,7} \foreach \c in {0,...,7} {%
    \pgfmathtruncatemacro\o{#3}\edef\oc{\own{\o}!#4}%
    \fill[\oc] ({#1+\c*\cs},{#2+(7-\r)*\cs}) rectangle ++(\cs,\cs);}
  \foreach \i in {1,...,7} {%
    \draw[white, line width=0.4pt] ({#1+\i*\cs},#2) -- ++(0,{8*\cs});
    \draw[white, line width=0.4pt] (#1,{#2+\i*\cs}) -- ++({8*\cs},0);}
  \draw[ink, line width=0.6pt] (#1,#2) rectangle ++({8*\cs},{8*\cs});}
\def\cell#1#2#3#4{({#1+(#4+0.5)*\cs},{#2+(7.5-#3)*\cs})}
\def\elem#1#2#3#4{%
  \draw[white, line width=2pt, line cap=round, rounded corners=2pt] #2;
  \draw[#1, line width=0.75pt, rounded corners=2pt] #2;
  \foreach \p in {#3,#4} {\fill[#1] \p circle (0.05); \draw[white, line width=0.4pt] \p circle (0.05);}}
\def\tick#1{\draw[ink, line width=0.8pt, line cap=round, line join=round, shift={(#1.east)}]
  (-0.17,0) -- (-0.125,-0.05) -- (-0.04,0.07);}
\def\cross#1{\draw[mute, line width=0.7pt, line cap=round, shift={(#1.east)}]
  (-0.155,-0.05) -- (-0.055,0.05) (-0.155,0.05) -- (-0.055,-0.05);}

\node[hd, anchor=base west] at (0,5.2) {Placement};
\node[sub, anchor=base west] at (0,4.95) {color: thread holding it};
\node[hd, anchor=base] at (4.1,5.2) {Route};
\node[sub, anchor=base] at (4.1,4.95) {cheapest that checks};
\node[hd, anchor=base east] at (8.2,5.2) {Demand};
\node[sub, anchor=base east] at (8.2,4.95) {color: thread wanting it};
\draw[mute!50, line width=0.5pt] (0,4.8) -- (8.2,4.8);

\node[plat] at (0,4.52) {B300\ \textmd{\color{mute}· tensor memory}};
\node[note, anchor=base east] at (8.2,4.52) {either consumer};
\grid{0}{3.22}{floor(\r/2)}{30}
\grid{7.0}{3.22}{floor(\r/2)}{100}
\elem{att}{\cell{0}{3.22}{1}{6} -- \cell{7}{3.22}{1}{6}}{\cell{0}{3.22}{1}{6}}{\cell{7}{3.22}{1}{6}}
\elem{com}{\cell{0}{3.22}{5}{1} -- \cell{7}{3.22}{5}{1}}{\cell{0}{3.22}{5}{1}}{\cell{7}{3.22}{5}{1}}
\node[rt] (r1) at (4.1,3.895) {tensor-memory read\kern0.25cm};
\tick{r1}

\node[plat] at (0,2.84) {H200, RTX 5090\ \textmd{\color{mute}· registers}};
\node[note, anchor=base east] at (8.2,2.84) {TileLang consumer};
\node[note, anchor=base east] at (8.2,1.3) {CUDA consumer};
\grid{0}{1.52}{floor(\c/2)}{100}
\grid{7.0}{1.52}{floor(\c/2)}{100}
\grid{7.0}{0}{floor(\r/2)}{100}
\def\scratch{(3.25,0.38) rectangle (4.95,0.82)}
\elem{com}{\cell{0}{1.52}{2}{5} -- (1.45,2.345) -- (3.3,0.7)}{\cell{0}{1.52}{2}{5}}{(3.3,0.7)}
\elem{att}{\cell{0}{1.52}{4}{1} -- (1.45,2.045) -- (3.3,0.5)}{\cell{0}{1.52}{4}{1}}{(3.3,0.5)}
\elem{cpl}{(4.9,0.7) -- (6.9,0.825) -- \cell{7}{0}{2}{5}}{(4.9,0.7)}{\cell{7}{0}{2}{5}}
\elem{com}{(4.9,0.5) -- (6.9,0.525) -- \cell{7}{0}{4}{1}}{(4.9,0.5)}{\cell{7}{0}{4}{1}}
\fill[ink!6] \scratch;
\fill[pattern=north east lines, pattern color=mute!45] \scratch;
\draw[ink, line width=0.7pt] \scratch;
\node[font=\sffamily\fX\bfseries, text=ink, fill=ink!6, inner sep=1pt] at (4.1,0.6) {shared scratch};
\elem{cpl}{\cell{0}{1.52}{1}{3} -- \cell{7}{1.52}{1}{3}}{\cell{0}{1.52}{1}{3}}{\cell{7}{1.52}{1}{3}}
\elem{com}{\cell{0}{1.52}{5}{4} -- \cell{7}{1.52}{5}{4}}{\cell{0}{1.52}{5}{4}}{\cell{7}{1.52}{5}{4}}
\node[rt] (r2) at (4.1,2.195) {register copy\kern0.25cm};
\tick{r2}
\node[font=\sffamily\fX, text=mute, fill=white, inner sep=1.2pt, anchor=west] (r3) at (2.75,0.16) {direct copy\kern0.25cm};
\cross{r3}
\node[rt, anchor=west] (r4) at ([xshift=0.1cm]r3.east) {store / load\kern0.25cm};
\tick{r4}
\end{tikzpicture}
\caption{Generated tensor transfers on the GEMM + SwiGLU case study
(schematic). Lines follow single elements from the thread that holds them to the
thread that wants them. The CUDA consumer wants elements in other threads, so the
direct copy fails the check and the linker stages through shared scratch.}
\label{fig:transfers}
\end{figure}

Instead, the producer's export publishes the tile's \emph{placement}: where
each element lives, as a tensor-memory lane and column, a shared-memory
address, or a thread and register. The consumer's import states its
\emph{demand}: which element each of its threads wants
(Figure~\ref{fig:transfers}). Both are exact maps fitted at pack time. The
linker then tries \emph{routes} from cheapest to most general: a read from
tensor memory straight into the demanded registers, a direct register copy
when placement and demand coincide, a warp-local shuffle, or staging through
scratch shared memory that the consumer lends. It checks each candidate against a model of the
hardware primitives for exact delivery of every element, freedom from races and
correct ordering, and lowers the first route that passes to CUDA through
TileLang~\cite{wang2026tilelang}.
Transfers move values exactly and never convert them, and their scratch is an
ordinary claim.

Two simpler designs were available. Always staging through shared memory works
for every pair, but it costs a round trip that the direct routes avoid
(Section~\ref{sec:eval-gemm-swiglu}). Fixing one exchange layout
by convention would make every producer and consumer agree in advance, which is
the coupling contracts remove. At the other extreme, general program synthesis
could find cleverer routes, but its results are hard to bound and to check. A
short ladder of route shapes, each checked exactly, covers every connection in
our case studies.

\subsection{Communication and Host Interface}
\label{sec:communication}
\label{sec:host}

Communication stays in component code. A component that moves data between
GPUs declares the communication domains it joins, the views it needs of a
symmetric allocation the caller lends (peer addresses, a multicast address),
and system-scope counters for readiness across GPUs. The linker checks that all
owners of an allocation agree on its domain and views, binds counter slots
like any other claim, and passes the caller's addresses to the owners. It does
not synthesize collectives: communication protocols are specific to their
algorithms (a ring push, a dispatch with routing, a reduce-scatter), and the
owners who design them are best placed to implement them. The linker's job is
to make such a component a well-behaved neighbor inside a shared kernel.

Host-side setup, such as building TMA descriptors, ships as a small compiled
provider behind a C interface. When a session opens, each owner's provider
fills its own region of the kernel's parameters from caller-lent buffers;
facts that other owners need flow through the manifest, never through shared
headers. An open session launches without host computation and can be captured
in a CUDA graph.

\section{Evaluation}
\label{sec:eval}

We evaluate PipeLink on four case studies (Table~\ref{tab:summary}). Each one
asks what a composition makes possible, and each also checks the cost of the
composition against the same component sources compiled as a single unit.

\subsection{Setup}
\label{sec:eval-setup}

\paragraph{Implementation.} The PipeLink compiler is about 24,000 lines of
Python. It packs components with \texttt{nvcc}, solves placement with Z3~\cite{demoura2008z3},
lowers generated transfers through TileLang~\cite{wang2026tilelang}, and links with nvJitLink~\cite{sundaram2023nvjitlink} from
CUDA~13.3. A TileLang frontend of about 5,400 lines derives contracts for
components written in TileLang. Code and reproduction scripts:
\url{https://github.com/PKU-DAIR/PipeLink}.

\paragraph{Hardware.} We use NVIDIA H200 (Hopper, sm90), B300 (Blackwell,
sm103) and RTX~5090 (consumer Blackwell, sm120) GPUs. Each machine runs every
timed experiment at one locked SM clock, the highest clock it sustains under
simultaneous dense BF16 and FP8 matrix multiplication on all GPUs: 1230\,MHz
on H200, 1155\,MHz on B300 and 1770\,MHz on RTX~5090.

\paragraph{Methodology.} For every linked kernel we build a \emph{single-unit
control} that includes the same component sources, the same generated
connecting code and the same placement in one translation unit, compiled with
the same floating-point mode. Linked and control images run in one process on
the same inputs, alternating in balanced blocks with a module reload halfway.
Every ratio in this section is a throughput ratio or speedup of the linked
kernel, so higher is better. Every linked kernel matches its control (bitwise,
or within run-to-run noise for the megakernels) and is free of memcheck,
racecheck and synccheck findings.

\begin{table}[t]
\caption{Case studies. \emph{Comp.}\ counts distinct component packages; a component
is \emph{reused} if more than one variant links it unchanged.}
\label{tab:summary}
\centering
\footnotesize
\setlength{\tabcolsep}{3pt}
\begin{tabular}{@{}llccc@{}}
\toprule
\textbf{Case study} & \textbf{Arch.} & \textbf{Comp.} & \textbf{Reused} & \textbf{Variants} \\
\midrule
Megakernel & sm103 & 10 & 8 & 2 \\
MoE megakernel & sm103 & 6 & 6 & 24 \\
Attention & sm90 & 14 & 8 & 7 \\
GEMM + SwiGLU & sm90/103/120 & 9 & 3 & 6 \\
\bottomrule
\end{tabular}
\end{table}

Across the 51 timed configurations, linked kernels run within 0.2\% of their
single-unit controls. For all seven attention
compositions, the mixed-library megakernel, and the GEMM + SwiGLU kernels on
H200 and RTX~5090, the normalized machine code of linked and control kernels
is identical. The case studies
therefore measure what composition enables, not what it costs.

\subsection{Cross-Library Megakernel}
\label{sec:eval-megakernel}

\paragraph{Composition.} We split the Llama-1B~\cite{meta2024llama} decode megakernel of Hazy
Research~\cite{spector2025megakernel}, written with ThunderKittens~\cite{spector2025thunderkittens}, into nine packages: an instruction
interpreter with five roles and eight operators.
The interpreter runs a precomputed instruction stream for 16 layers in one
persistent kernel; operators claim shared-memory pages, semaphores and scratch
for the duration of their instruction phase, so different operators reuse the
same pool. The original megakernel fails our correctness checks: its split-KV
path deadlocks or never stores its partial outputs, and racecheck reports
hazards where warps hand shared memory to each other. We repaired five upstream
defects in every image, the monolith included. Linked, the all-ThunderKittens kernel reaches
0.9995--1.0002$\times$ the throughput of its single-unit control across five
configurations, and 1.01--1.10$\times$ the throughput of the repaired monolith.

\paragraph{Cross-library replacement.} We then replace only the attention-partial
package with one written in CUTLASS/CuTe~\cite{thakkar2023cutlass}: four warps process disjoint key/value
tiles with warp-level MMA and merge their normalized partial results within the
component. The other eight packages, the instruction stream and the compiled host
provider are unchanged; the CuTe package satisfies the same contract, so the
composition links without any change to the interpreter.
Table~\ref{tab:megakernel} shows the result. At a 4096-token context, the
mixed-library kernel decodes 2.16$\times$ faster than the ThunderKittens
kernel; at 1024 tokens with split-KV reduction, 1.22$\times$ faster. At a
16-token context it is 5\% slower, because one of the four warps has work and
the component's fixed copy and merge costs dominate. The mixed kernel reaches
0.9994--0.9998$\times$ the throughput of its own single-unit control and has zero sanitizer findings.

\begin{table}[t]
\caption{Llama-1B decode megakernel on B300 (1155\,MHz), ms per token; TK is ThunderKittens.}
\label{tab:megakernel}
\centering
\small
\begin{tabular}{@{}lrrr@{}}
\toprule
\textbf{Context} & \textbf{TK only} & \textbf{TK + CuTe} & \textbf{Speedup} \\
\midrule
16 & 0.713 & 0.752 & 0.95$\times$ \\
1024, reduction & 1.151 & 0.939 & 1.22$\times$ \\
4096 & 3.516 & 1.630 & 2.16$\times$ \\
\bottomrule
\end{tabular}
\end{table}

\subsection{Multi-Owner MoE Megakernel}
\label{sec:eval-moe}

\paragraph{Composition.} Mixture-of-Kittens~\cite{sul2026mixture} is a deterministic expert-parallel
MoE training megakernel for Blackwell.
We split it among four owners: a framework that owns the task interpreter, the
cluster launch control and the launch geometry, and three operator owners for
communication (dispatch and combine over NVLink), grouped GEMM, and SwiGLU
activation. Each owner builds its own descriptors from caller-lent buffers in a
compiled host provider; tile geometry that another owner schedules against is a
declared fact, and synchronization between owners runs through counter channels,
some with three producers. The framework provides one interpreter package each
for the forward, backward and recompute passes; each operator owner ships a
single package whose exports serve all three passes and all variants (BF16 and
MXFP8, one or four GPUs, clamped or not). These six packages link into all 24
kernels.

\paragraph{Performance.} On four B300 GPUs at 2048 and 8192 tokens per rank, the
12 timed kernels reach 0.999--1.001$\times$ the throughput of their single-unit
controls and 0.989--1.014$\times$ the throughput of the original monolithic
MoK step, measured in the same processes with the same inputs and routing.

\paragraph{Owner replacement.} An independent activation owner implements the
GPT-OSS~\cite{openai2025gpt} variant of SwiGLU against the same contract. It is built in a source
tree that contains none of the other owners' implementation sources, links into
all 24 kernels with the other owners' packages and providers unchanged, and
passes all 48 numerical and bitwise checks per rank.

\paragraph{Owner upgrades.} All five single-constant upgrades of
Section~\ref{sec:symbol-linking}, which hang or leave outputs unwritten when
linked by symbol, relink correctly: PipeLink propagates each new fact to the owners that
consume it. Seven incompatible combinations, such as a BF16-only communication
package in an MXFP8 kernel, are each rejected with a named diagnostic.

\subsection{Attention Variants from One Core}
\label{sec:eval-attention}

\begin{figure}[t]
\centering
\def\fX{\fontsize{6}{7}\selectfont}\def\fS{\fontsize{7}{8.4}\selectfont}\def\fT{\fontsize{8.5}{10}\selectfont}%
\begin{tikzpicture}[
  x=1cm, y=1cm, font=\sffamily\fS,
  hd/.style={font=\sffamily\fX\bfseries, text=mute, anchor=base, inner sep=0},
  row/.style={font=\sffamily\fS, text=ink, anchor=east, inner sep=0},
]
\definecolor{att}{HTML}{2B6CB0}
\definecolor{cpl}{HTML}{2F855A}
\definecolor{com}{HTML}{C05621}
\definecolor{ink}{HTML}{1A202C}
\definecolor{mute}{HTML}{718096}

\def\cwid{0.86}\def\rh{0.34}
\def\at#1#2{\pgfmathsetmacro\X{1.95+#1*0.94}\pgfmathsetmacro\Y{-#2*0.44-\rh/2}}
\def\dflt#1#2{\at{#1}{#2}\fill[mute!16, rounded corners=1.5pt] (\X,\Y) rectangle ++(\cwid,\rh);}
\def\idle#1#2{\at{#1}{#2}\draw[mute!70, densely dashed, line width=0.5pt, rounded corners=1.5pt]
  (\X,\Y) rectangle ++(\cwid,\rh);}
\def\new#1#2#3#4{\at{#1}{#2}%
  \fill[#3!12, rounded corners=1.5pt] (\X,\Y) rectangle ++(\cwid,\rh);
  \draw[#3, line width=0.8pt, rounded corners=1.5pt] (\X,\Y) rectangle ++(\cwid,\rh);
  \node[font=\sffamily\fX\bfseries, text=#3, inner sep=0] at ({\X+\cwid/2},{\Y+\rh/2}) {#4};}

\foreach \c/\t in {0/load, 1/core, 2/mask, 3/score, 4/epilogue, 5/comm.}
  \node[hd] at ({1.95+\c*0.94+\cwid/2},0.32) {\t};
\draw[mute!50, line width=0.5pt] (0,0.25) -- ({1.95+5*0.94+\cwid},0.25);

\foreach \r/\t in {0/causal, 1/window, 2/softcap, 3/window + softcap, 4/Phi-3 block-sparse,
                   5/MSA, 6/context parallel}
  \node[row] at (1.85,{-\r*0.44}) {\t};

\foreach \r in {0,1,2,3} {\dflt{0}{\r}}
\new{0}{4}{cpl}{block}
\new{0}{5}{cpl}{sparse}
\new{0}{6}{cpl}{CP}
\foreach \r in {0,2,5,6} {\dflt{2}{\r}}
\new{2}{1}{cpl}{window}
\new{2}{3}{cpl}{window}
\new{2}{4}{cpl}{block}
\foreach \r in {0,1,4,5,6} {\dflt{3}{\r}}
\new{3}{2}{cpl}{softcap}
\new{3}{3}{cpl}{softcap}
\foreach \r in {0,1,2,3,4,6} {\dflt{4}{\r}}
\new{4}{5}{cpl}{partial}
\foreach \r in {0,...,5} {\idle{5}{\r}}
\new{5}{6}{com}{push}

\def\core{(2.89,{-6*0.44-\rh/2}) rectangle ({2.89+\cwid},{\rh/2})}
\fill[att!10, rounded corners=2pt] \core;
\fill[pattern=north east lines, pattern color=att!28] \core;
\draw[att, line width=0.9pt, rounded corners=2pt] \core;
\node[font=\sffamily\fS\bfseries, text=att, fill=att!10, rotate=90, inner sep=1.5pt]
  at ({2.89+\cwid/2},{-3*0.44}) {same compiled package};

\def\ly{-3.2}
\fill[mute!16, rounded corners=1pt] (1.95,{\ly-0.08}) rectangle ++(0.3,0.16);
\node[font=\sffamily\fX, text=ink, anchor=west, inner sep=0] at (2.32,\ly) {default, reused};
\fill[cpl!12, rounded corners=1pt] (4.0,{\ly-0.08}) rectangle ++(0.3,0.16);
\draw[cpl, line width=0.7pt, rounded corners=1pt] (4.0,{\ly-0.08}) rectangle ++(0.3,0.16);
\node[font=\sffamily\fX, text=ink, anchor=west, inner sep=0] at (4.37,\ly) {new};
\draw[mute!70, densely dashed, line width=0.5pt, rounded corners=1pt] (5.1,{\ly-0.08}) rectangle ++(0.3,0.16);
\node[font=\sffamily\fX, text=ink, anchor=west, inner sep=0] at (5.47,\ly) {idle warps};
\end{tikzpicture}
\caption{Seven attention variants built around one unchanged core. Each row is a
linked kernel that links the same compiled core package; the orange cell is the
communication component.}
\label{fig:attention}
\end{figure}
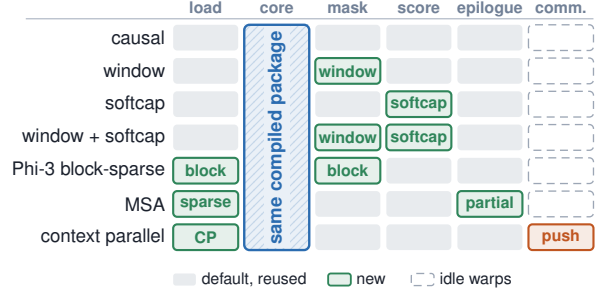

\paragraph{Composition.} We derived one attention core for Hopper from
FlashAttention-3~\cite{shah2024flashattention} (BF16, head dimension 128) and froze its package.
Seven compositions reuse it unchanged, each adding or replacing at most two
components (Figure~\ref{fig:attention}). Sliding windows and softcapping are
mask and score callables. Phi-3~\cite{microsoft2024phi} block-sparse attention and MiniMax sparse
attention (MSA)~\cite{lai2026minimax} replace the load component with one that visits only the
selected key blocks.
Context-parallel attention (CP) replaces the load component and adds
a communication component, which pushes key/value
blocks to the other GPUs and releases a readiness counter per block. Because
every query tile visits all of its key/value blocks inside one kernel, no
output correction is needed.
All 17 timed single-GPU configurations reach 0.9991--1.0005$\times$ the
throughput of their single-unit controls, and CP 0.9997$\times$ and
0.9992$\times$ on two and four GPUs.

\paragraph{Placement.} Table~\ref{tab:attention} compares the variants with
public implementations~\cite{shah2024flashattention,ye2025flashinfer} on released-model workloads. The sparse variants gain
most, because their load components skip unselected blocks: MSA is
1.6--4.5$\times$ faster than vLLM's~\cite{kwon2023efficient} dedicated kernel and up to 38$\times$
faster than FlexAttention~\cite{dong2025flexattention}. FA4~\cite{zadouri2026flashattention}, which masks the selection over query tiles, is
5--6\% faster when the selected blocks form a prefix but 9--38$\times$ slower
when they are scattered.

\begin{table}[t]
\caption{Attention placement on H200 (1230\,MHz): throughput of the linked
kernel relative to the best measured alternative. MSA uses one reference
for both selection patterns, the alternative fastest over both combined (FA4 is faster on prefixes only).}
\label{tab:attention}
\centering
\small
\begin{tabular}{@{}llr@{}}
\toprule
\textbf{Workload} & \textbf{Reference} & \textbf{Linked / ref.} \\
\midrule
Qwen3-8B causal, 8K--32K & FA3 & 0.96--0.99 \\
Mistral/Gemma sliding window & FA3 & 0.84--0.94 \\
Gemma2/Grok softcap & FA3 & 0.92--0.94 \\
Llama 4 chunked, 32K--128K & FA3 & 1.01--1.02 \\
Phi-3 block-sparse, 32K--128K & best expert & 2.26--2.73 \\
MSA, prefix selection & \multirow{2}{*}{vLLM Triton} & 4.47--4.52 \\
MSA, uniform selection & & 1.62--1.63 \\
\bottomrule
\end{tabular}
\end{table}

\paragraph{Context parallelism.} For causal prefill over 32K tokens, CP is
1.95$\times$ faster on two H200 GPUs and 3.67$\times$ faster on four than the
single-GPU kernel (97.6\% and 91.7\% of ideal). Against Transformer Engine's~\cite{nvidia2026transformer}
context parallelism on the same GPUs, using its faster transport and capturing
its forward pass in a CUDA graph, CP is 1.20--1.31$\times$ faster at 8K and
32K tokens (Table~\ref{tab:cp}). Both sides use their own local attention
kernels, so the comparison covers complete implementations.

\begin{table}[t]
\caption{Context-parallel causal attention on H200 (1230\,MHz), ms.}
\label{tab:cp}
\centering
\small
\begin{tabular}{@{}lrrrr@{}}
\toprule
\textbf{Tokens} & \textbf{GPUs} & \textbf{PipeLink} & \textbf{TE (p2p)} & \textbf{Speedup} \\
\midrule
8K & 2 & 0.604 & 0.750 & 1.24$\times$ \\
8K & 4 & 0.375 & 0.490 & 1.31$\times$ \\
32K & 2 & 7.795 & 9.446 & 1.21$\times$ \\
32K & 4 & 4.133 & 4.964 & 1.20$\times$ \\
\bottomrule
\end{tabular}
\end{table}

\subsection{Fused GEMM and Activation}
\label{sec:eval-gemm-swiglu}

\paragraph{Composition.} We compose the MLP up-projection of Qwen3-32B~\cite{qwenteam2025qwen} with its
SwiGLU activation: $X[M,5120]$ times interleaved gate and up weights
$W[51200,5120]$, producing the BF16 activation $[M,25600]$ for $M=16384$ and
4096. The FP32 accumulator stays unrounded through the activation. Splitting a
fused kernel into a GEMM producer and an elementwise consumer follows CODA~\cite{guo2026coda}; our
components and shapes differ.
Each architecture has its own GEMM producer: a two-SM tensor-core GEMM on B300
that leaves its accumulator in tensor memory, a three-warpgroup WGMMA GEMM on
H200 and a multistage \texttt{mma.sync} GEMM on RTX~5090, both leaving it in
registers. The two consumers of Section~\ref{sec:authoring}, one in CUDA and
one in TileLang, import the accumulator in 32-column chunks. The producers link
without their sources.

\paragraph{Generated transfers.} On B300 the linker reads the accumulator from
tensor memory into each consumer thread's registers. On H200 and RTX~5090 it
stages each chunk through 16\,KiB of shared memory for the CUDA consumer and
gives the specialized TileLang consumer a direct register copy
(Figure~\ref{fig:transfers}). On H200, the direct route makes the TileLang
kernel 5\% faster than the CUDA one (14.35 versus 15.10\,ms at $M=16384$).

\paragraph{Placement.} Table~\ref{tab:swiglu} compares the linked kernels with
public and native baselines. On B300, the composition is 1.20--1.21$\times$
faster than cuBLAS~\cite{nvidia2026cublas} followed by a \texttt{torch.compile}d~\cite{ansel2024pytorch} SwiGLU and
1.34--1.37$\times$ faster than cuBLAS followed by Liger's~\cite{hsu2024liger} SwiGLU. Against
fusing the same components by hand into one native kernel, the linked kernels
reach 0.94--1.01$\times$ its throughput. All 12 configurations reach
0.9987--1.0015$\times$ the throughput of their single-unit controls.

\begin{table}[t]
\caption{Qwen3-32B GEMM + SwiGLU: speedup of the linked kernel over each
baseline at $M=16384$; CUDA / TileLang consumer.}
\label{tab:swiglu}
\centering
\footnotesize
\setlength{\tabcolsep}{3pt}
\begin{tabular}{@{}lccc@{}}
\toprule
\textbf{Baseline} & \textbf{B300} & \textbf{H200} & \textbf{5090} \\
\midrule
cuBLAS + compiled & 1.20 / 1.20 & 1.03 / 1.08 & 1.03 / 1.01 \\
cuBLAS + Liger & 1.34 / 1.34 & 1.12 / 1.18 & 1.08 / 1.05 \\
Same GEMM, unfused & 1.11 / 1.10 & 1.02 / 1.07 & 1.03 / 1.01 \\
Hand-fused native & 0.99 / 0.98 & 0.94 / 0.99 & 1.01 / 0.99 \\
\bottomrule
\end{tabular}
\end{table}

\section{Related Work}
\label{sec:related}

We group prior work by where the information needed to compose kernel parts
lives (Table~\ref{tab:related}).

\begin{table}[t]
\caption{How approaches compose parts of a GPU kernel (WS: warp-specialization;
RTs: runtimes; res.: on-chip resources; comm.: cross-GPU communication inside
the kernel). \yes{}~yes, \partly{}~partly, \no{}~no.}
\label{tab:related}
\centering
\footnotesize
\setlength{\tabcolsep}{2.5pt}
\begin{tabular}{@{}lccccc@{}}
\toprule
& \shortstack{\textbf{Binary}\\\textbf{parts}} & \shortstack{\textbf{Cross-}\\\textbf{lib.}}
& \shortstack{\textbf{Places}\\\textbf{res.}} & \shortstack{\textbf{Checks}\\\textbf{iface.}}
& \shortstack{\textbf{Kernel}\\\textbf{comm.}} \\
\midrule
Templates, DSLs & \no & \no & \partly & \partly & \no \\
WS compilers & \no & \no & \yes & \yes & \no \\
Megakernel RTs & \no & \no & \partly & \partly & \partly \\
Symbol linking & \yes & \yes & \no & \no & \no \\
Comm.\ overlap & \no & \no & \no & \no & \yes \\
\textbf{PipeLink} & \yes & \yes & \yes & \yes & \yes \\
\bottomrule
\end{tabular}
\end{table}

\paragraph{In one source program.} Template libraries such as
CUTLASS~\cite{thakkar2023cutlass} and attention libraries that accept mask and
score functions or generate variants from templates~\cite{dong2025flexattention,ye2025flashinfer,chen2025attentionengine,you2025flashlight,chen2026metaattention}
compose at extension points that their authors designed. Domain-specific
languages such as TileLang~\cite{wang2026tilelang}, CuTe DSL~\cite{nvidia2026cutedsl},
ThunderKittens~\cite{spector2025thunderkittens} and Triton~\cite{tillet2019triton},
and fusion libraries such as FKL~\cite{amoros2025fused} and
CODA~\cite{guo2026coda}, compose within one program; sparse-attention kernels
specialize execution to the selection pattern in the same
way~\cite{zhao2026quilt,li2026pageweaver,liu2026decoupling}. PipeLink instead
links components that different authors compiled separately, and can link
components written with these tools.

\paragraph{In compiler IR.} Warp-specialization compilers partition a kernel
into roles and synchronize them: Tawa~\cite{chen2026tawa} uses asynchronous
references as channels, Twill~\cite{soi2025optimal} solves pipelining and warp
specialization jointly, Cypress~\cite{yadav2025task} derives data movement from
a mapping, and TLX~\cite{guan2026tlx} and Gluon~\cite{triton2026gluon} expose
roles and memory to the programmer. Layout
systems~\cite{zhang2026hexcute,zhou2026linear,hou2026axe} represent the layouts
that connect operations. These compilers own and may reschedule the whole
kernel; PipeLink keeps each component's compiled schedule and works over
declared interfaces of opaque objects.

\paragraph{In runtime task graphs.} Megakernel systems schedule operators or
tiles across SMs in one persistent kernel: MPK~\cite{cheng2025mirage} and Event
Tensor~\cite{jin2026event} compile models into in-kernel task graphs, and other
work tunes, searches or generates megakernel schedules, also with coding
agents~\cite{spector2025megakernel,he2026taming,he2026vdcores,jaber2026automegakernel,dong2026ada,chowdhary2026fleet,li2026forgemegakernel,yao2026megaflux}.
These systems decide which task runs where; PipeLink decides how separately
compiled components share one SM, so their task bodies could be linked
components, as in our Llama-1B megakernel.

\paragraph{By symbol.} Relocatable device code and LTO~\cite{scudiero2014separate},
exposed at run time by nvJitLink~\cite{sundaram2023nvjitlink}, link separately
compiled device functions at no runtime cost; cuFFT~\cite{nvidia2026cufftlto}
and cuFFTDx~\cite{nvidia2026cufftdx} use them to fuse callbacks and library code,
and GoPTX~\cite{wu2025goptx} fuses kernels at the PTX level. PipeLink adds what
concurrently running components share and a linker that checks and places it.

\paragraph{Communication overlap.} Many systems overlap communication with
computation through streams, signals or fused
kernels~\cite{chang2024flux,zheng2025tilelink,zheng2025distributed,awad2025iris,sul2025parallelkittens,qiang2026syncopate,mao2026mkernel,ma2026entwine,hong2026overlapping,gond2025tokenweave,zhang2025comet,nvidia2026nvshmem},
and context-parallel frameworks correct outputs outside the attention
kernel~\cite{nvidia2026transformer,fang2024usp,gu2024loongtrain,yang2025context}.
PipeLink contributes no communication algorithm; it lets one owner's
communication component join another owner's compute components in one kernel.

\paragraph{Synchronization verification.} WEFT~\cite{sharma2015verification},
GPUVerify~\cite{betts2012gpuverify}, Descend~\cite{kopcke2024descend} and
Exo-GPU~\cite{akeley2026exo} verify synchronization or schedule safety of whole
programs. PipeLink checks declared protocols between components and leaves
their code to sanitizers and numerical tests.

\section{Limitations and Conclusion}
\label{sec:conclusion}

\paragraph{Limitations.} PipeLink checks contracts against each other, not
against the code behind them. A component that breaks its own contract, for
example by touching shared memory outside its claim, can still produce a wrong
kernel; numerical checks and sanitizers therefore remain part of every
composition, as they are for any hand-fused kernel.

The code PipeLink generates only moves data, and only within one thread
block. Arithmetic, precision conversion and communication between thread blocks
or GPUs stay in component code, where the linker checks and binds the counters
and memory they use but does not write them. A tensor export carries one
generation, so a producer that hands its accumulator to another component in
this way cannot be persistent; multi-slot rings between components are written
in component code.

PipeLink does not make components faster. It preserves each owner's schedule,
so a composition performs as well as the components it links; the gap of our
dense attention variants to FlashAttention-3 comes from our core. Existing
kernels must also be ported to declared interfaces; the four-owner MoE
megakernel shows what such a port involves. Finally, packages are compiled for
one architecture and one toolkit, because LTO IR is not portable across CUDA
major versions.

\paragraph{Conclusion.} The components of fused GPU kernels cooperate at fine
grain on one SM, yet today they are absolute-addressed and can only be combined
by forking their source. PipeLink gives them what object files gave whole
programs: a contract stating what crosses the component boundary, and a linker
that checks the contracts and relocates every component into the kernel it
joins. Because the linker's decisions fold into constants, composition costs
nothing, and components from different libraries, owners and GPUs combine into
kernels that were previously built by hand or not at all.

\bibliography{references}
\bibliographystyle{mlsys2025}

\end{document}